\documentclass[aps,prl,reprint,groupedaddress,nofootinbib,superscriptaddress]{revtex4-1}

\begin{document}

\preprint{RESCEU9/19}

\title{Horndeski model in nonlinearly realized supergravity}


\author{Yusuke Yamada}
\email{yamada@resceu.s.u-tokyo.ac.jp}
\affiliation{Research Center for the Early Universe (RESCEU), Graduate School of Science, The University of Tokyo, Hongo 7-3-1
Bunkyo-ku, Tokyo 113-0033, Japan}
\author{Jun'ichi Yokoyama}
\email{yokoyama@resceu.s.u-tokyo.ac.jp}
\affiliation{Research Center for the Early Universe (RESCEU), Graduate School of Science, The University of Tokyo, Hongo 7-3-1
Bunkyo-ku, Tokyo 113-0033, Japan}
\affiliation{Department of Physics, Graduate School of Science,
The University of Tokyo, Hongo 7-3-1
Bunkyo-ku, Tokyo 113-0033, Japan}
\affiliation{Kavli Institute for the Physics and Mathematics of the Universe (Kavli IPMU),
UTIAS, WPI, The University of Tokyo, Kashiwa, Chiba, 277-8568, Japan}


\date{\today}

\begin{abstract}
We construct the Horndeski Lagrangian within non-linearly realized supergravity. We will show that the bosonic part of the Horndeski Lagrangian can be realized. Gravitino naturally couples to Horndeski sector in a super-covariant way. Such gravitino couplings are also free from ghosts.
\end{abstract}


\maketitle

%
One of the most important goals of contemporary physics is to construct a theory to describe the entire
history of the Universe consistent with observations
from its creation to the current state.  
One may expect the birth of the classical Universe is 
 described by string theory which is the most promissing
candidate of quantum gravity.  Then physics of inflation, which is an indispensable ingredient of modern cosmology
to solve the horizon and the flatness problems as well as to generate curvature perturbations \cite{Sato:2015dga}, 
is plausibly described
by supergravity as the low energy effective theory of superstring, 
since observational
constraints on the tensor perturbations tell us 
that the energy scale of inflation is at least several orders smaller than
the Planck scale (at that time) \cite{Ade:2018gkx}.

%
In order to realize inflation in supergravity, first of all, we must realize a (quasi)  de Sitter universe which inevitably breaks
supersymmetry.
Recently the simplest supergravity model of de Sitter space, dubbed pure
de Sitter supergravity, was
proposed~\cite{Bergshoeff:2015tra,Hasegawa:2015bza}, in which the
physical degrees of freedom are graviton and massive gravitino. In this
model, spontaneous supersymmetry breaking is realized with a constrained
chiral superfield $S(x,\theta)$ satisfying the nilpotent condition
$S^2(x,\theta)=0$~\cite{Rocek:1978nb,Ivanov:1978mx,Lindstrom:1979kq},
which describes the Goldstino first found by Volkov and
Akulov~\cite{Volkov:1973ix}. Such a superfield is also known to be the
low energy effective description of anti-D3 brane in string
theory~\cite{Kallosh:2014wsa, Bergshoeff:2015jxa,
Kallosh:2015nia,Garcia-Etxebarria:2015lif,Cribiori:2019hod}. The
nilpotent superfield has also been applied  to supergravity cosmology models e.g. in~\cite{Antoniadis:2014oya,Ferrara:2014kva,DallAgata:2014qsj,McDonough:2016der,Kallosh:2017wnt}.

Turning our eyes to the bottom-up approach to inflationary cosmology,
 the cosmic microwave background (CMB) observations strongly favor
 single-field inflation as we do not find any trace of isocurvature
 modes \cite{Akrami:2018odb}.  Then the generalized G-inflation \cite{Kobayashi:2011nu} is the theory we should consider because it is the most general single-field inflation model containing practically all the known models of single-field inflation from potential driven models including the Starobinsky model \cite{Starobinsky:1980te}
 to kinetically driven ones
 \cite{ArmendarizPicon:1999rj,Kobayashi:2010cm} with or without various
 nonminimal couplings to gravity \cite{Spokoiny:1984bd,Germani:2010gm}.
This model is based on the Horndeski theory \cite{Horndeski:1974wa}
or the generalized galileon \cite{Deffayet:2011gz} which are the most general covariant scalar-tensor
theory whose field equations are of second-order so that ghost instability is absent.


It is unanimously agreed that sensible theories should not contain
fundamental ghost fields
\cite{Carroll:2003st,Woodard:2006nt,Sbisa:2014pzo}, but ghost-free
nature is important even from low energy effective field theory (LEFT) viewpoint.
In general, LEFT contains infinite number of derivative interactions originating from integrating heavy fields out, which would be schematically given by $f(\frac{\partial}{M_H}, \phi_L)$ where $M_H$ denotes heavy field mass and $\phi_L$ being light fields.  For instance, the propagator of a heavy scalar field  $M_H^{-2}(1+\partial^2/M_H^2)^{-1}$ gives rise to such interaction.  All the terms are suppressed by heavy field mass and $M_H$ becomes the cut-off scale of LEFT. 
On the other hand, ghost-free type higher-derivative terms does not have
any pole of heavy fields.  
Therefore, even if they dominate the dynamics, any heavy fields would
not be excited, and ghost-free higher derivative terms can be the dominant part of dynamics 
as long as strong coupling regime is avoided.  
In other words, among various higher-derivative terms, 
only ghost-free ones can be leading order terms. 
Also, the ghost-free type interactions may have 
hidden symmetry~\cite{Aoki:2018lwx,Aoki:2019rvi}, 
which becomes manifest in the affine-metric formulation.

The purpose of this work is to construct the Horndeski model coupled to
pure de Sitter supergravity, namely, to embed the Horndeski Lagrangian
within nonlinearly realized supergravity to make the generalized
G-inflation possible in supergravity\footnote{One might wonder 
if this is a marriage of losers, as LHC does not discover supersymmetry yet, and trivial but strong constraints have been obtained against the Horndeski 
theory from GW170817 and GRB 170817A in terms of the propagation speed of gravitational waves \cite{Baker:2017hug,Creminelli:2017sry,Sakstein:2017xjx,Ezquiaga:2017ekz} using the formula derived in \cite{Kobayashi:2011nu}.  (See also \cite{Copeland:2018yuh} for nontrivial analysis.)  These arguments only apply in the low energy regime well below TeV scale and we should consider the full theories to describe physics of inflation \cite{Kobayashi:2019hrl}.}.
Supergravity realization of ghost free higher-derivative interaction has been a difficult issue, since supersymmetry predicts not only the desired term but also additional derivative couplings, which lead to ghost instabilities. Ghost-free higher derivatives in supersymmetric models are discussed e.g. in~\cite{Khoury:2010gb,Khoury:2011da,Farakos:2012je,Sasaki:2012ka,Koehn:2012ar,Koehn:2012np,Farakos:2012qu,Farakos:2013zya,Gwyn:2014wna,Aoki:2014pna,Aoki:2015eba,Fujimori:2016udq,Fujimori:2017kyi,Fujimori:2017rcc,Nitta:2018yzb,Nitta:2018vyc}. As we will show, all the issues of ghost instabilities are circumvented once supersymmetry is nonlinearly realized, and the Horndeski Lagrangian is consistently embedded within supergravity.
In particular, the resultant supersymmetric Horndeski model has couplings to gravitino without ghost instabilities.


Let us start with a review the construction of the pure de Sitter supergravity model~\cite{Bergshoeff:2015tra,Hasegawa:2015bza}. Throughout the manuscript, we  use the notation of~\cite{Wess:1992cp}.

The pure de Sitter supergravity model consists of the standard supergravity multiplet and a nilpotent chiral superfield $S(x,\theta)$~\cite{Rocek:1978nb,Ivanov:1978mx,Lindstrom:1979kq} satisfying 
\begin{equation}
S^2(x,\theta)=0.
\end{equation}
The nontrivial solution to this constraint is that the lowest component of $S(x,\theta)|_{\theta=0}=s(x)$ is no longer an independent degree of freedom, but becomes the fermion bilinear
\begin{equation}
s(x)=\frac{\chi_S^\alpha \chi_{S\alpha}}{2F^S},\label{S}
\end{equation}
where $\chi_S$ 
is the left handed Weyl spinor and $F^S$ the auxiliary component in
$S(x,\theta)$. Since the auxiliary field $F^S$ is in the denominator
in~(\ref{S}), the expectation value of $F^S$ should be non-vanishing
everywhere. In other words, this constrained superfield can be defined
only when $\langle F^S\rangle\neq 0$, namely supersymmetry is
spontaneously broken. 
This is not a problem as the supersymmetry is broken in de Sitter space anyway.
One may think of this constrained superfield as an effective description of sGoldstino $s(x)$ decoupling~\cite{Komargodski:2009rz}.

Due to the constraint on $S(x,\theta)$, the most general superpotential is 
\begin{equation}
W=W_0+\mu^2 S,\label{W}
\end{equation}
where $W_0$ and $\mu$ are real constants. Besides that, the form of the K\"ahler potential is restricted to
\begin{equation}
K=S\bar{S}.\label{K}
\end{equation}
The action of the pure de Sitter supergravity is given by
\begin{equation}
\mathcal{L}=\int d^2 \Theta 2{\cal E}\left[\frac{3}{8}\left(\bar{\cal D}^2-8{\cal R}\right)e^{-\frac{K}{3}}+W\right]+{\rm h.c.}\label{PSG}
\end{equation}
where $W$ and $K$ are defined in (\ref{W}) and (\ref{K}). One needs to integrate out the auxiliary complex scalars $F^S$, $M$ and vector $b_a$, which requires straightforward but tedious calculations. The complete component action including fermions is shown in~\cite{Bergshoeff:2015tra,Hasegawa:2015bza}. Here we will focus on the system in the unitary gauge of supersymmetry:
\begin{equation}
\frac{1}{\sqrt{2}}{\cal D}^\alpha S|_{\theta=0}=\chi_S^\alpha=0,\label{unitary}
\end{equation}
which significantly simplifies the Lagrangian. Under the unitary gauge condition, the system is simply given by the standard supergravity action with $\chi_S=S=0$, 
\begin{eqnarray}
\mathcal{L}=&&-e\Biggl[\frac{1}{2}R+\varepsilon^{klmn}\bar{\psi}_k\bar{\sigma}_l\tilde{\cal D}_m\psi_n\nonumber\\
&&-W_0(\psi_a\sigma^{ab}\psi_b+{\rm h.c.})-(\mu^2-3W_0^2)\Biggr].
\end{eqnarray}
Here $R$ is the Ricci scalar and $\psi_\mu$ is the gravitino. Here, the
physical degrees of freedom are graviton $e_{\mu}^a$ and massive
gravitino $\psi_\mu$ and nothing else. The mass of gravitino $m_{3/2}$
is given by $m_{3/2}=W_0$. The constant part
$\Lambda^4\equiv\mu^2-3W_0^2$ corresponds to vacuum energy density. 
Note that if $\mu=0$ the cosmological constant can never be positive,
which was the standard lore in supergravity: the pure supergravity
system with graviton and gravitino can only lead to negative
cosmological constant, namely, anti-de Sitter spacetime. 
However, with a nilpotent superfield, we can realize the system having graviton, massive gravitino and positive cosmological constant, which is the pure de Sitter supergravity.


We now show the Horndeski model~\cite{Horndeski:1974wa,Deffayet:2011gz,Kobayashi:2011nu} coupled to pure de Sitter supergravity. In the case with linearly realized supersymmetry, the couplings having derivatives more than two lead to ghosts or dynamically propagating auxiliary fields, which are obstacles to realize Horndeski type interactions. So far, the known ghost free higher-derivative interactions with linearly realized supergravity are only of the kind of $\mathcal{L}_2=P(X,\phi)$~\cite{Khoury:2010gb,Koehn:2012ar} and $\mathcal{L}_5=G^{\mu\nu}\partial_\mu\phi \partial_{\nu}\phi$~\cite{Farakos:2012je} where $X=-\frac12\partial_\mu\phi\partial^\mu\phi$ and $G_{\mu\nu}$ is the Einstein tensor. The latter is only known in the new minimal supergravity, and more general coupling ${\tilde G}_{5}(X,\phi)G^{\mu\nu}\partial_\mu\phi \partial_\nu\phi$ without ghosts has not been found yet.

The situation is drastically different in the nonlinearly realized
supersymmetry as we will show. It seems rather natural to expect that
Horndeski type interaction can be realized if supersymmetry is
spontaneously broken:  When supersymmetry is broken at some scale, its
very low energy effective theory, where the heavy superparticles 
decouple, would become effectively non-supersymmetric system, and Horndeski type Lagrangian can be realized in such case. As pointed out in~\cite{Farakos:2017mwd}, the following expression would realize (almost) arbitrary couplings within supergravity:
\begin{eqnarray}
&&\int d^4\theta E \frac{16 S\bar{S}}{{\cal D}^\alpha{\cal D}_\alpha S \bar{\cal D}_{\dot\alpha}\bar{\cal D}^{\dot\alpha}\bar{S}} {\cal F}({\cal C}, {\cal D}_\alpha {\cal C}, {\cal D}_a{\cal C})\nonumber\\
&&={\cal F}(C, \zeta_\alpha, \nabla_a C)+\cdots,
\label{general}
\end{eqnarray}
where ${\cal C}$ is a general superfield, which may also have spinor or vector indices in general, and $C={\cal C}|_{\theta=0}$. We have also defined $\zeta_\alpha\equiv {\cal D}_\alpha{\cal C}|_{\theta=0}$. The derivative ${\cal D}_a$ is vector derivative in superspace and $\nabla_a=e_a^{m}\nabla_m$ denotes the spacetime derivative covariant under diffeomorphism as well as supersymmetry. The ellipses part denotes the terms including the Goldstino $\chi_S$, which vanish in the unitary gauge~(\ref{unitary}).

The coupling~(\ref{general}) is crucial to realize the Horndeski model. From the above observation, it would be enough to find the superfield function $\mathcal{F}$ which has the Horndeski couplings in the lowest component. Since the original Horndeski model is described by a single real scalar field, we consider a chiral superfield $\Phi$ satisfying the following constraint~\cite{Komargodski:2009rz,Farakos:2017mwd,Ferrara:2015tyn,Carrasco:2015iij,DallAgata:2015zxp,Ferrara:2016een}
\begin{equation}
S(\Phi-\bar{\Phi})=0,\label{const}
\end{equation}
where $S$ is the nilpotent superfield. All, but ${\rm Re}\Phi|$, 
of components in $\Phi$ become dependent fields of $\chi_S^\alpha$ 
and vanish in unitary gauge, namely
\begin{equation}
\Phi(x,\theta)=\phi(x) \ \text{(in unitary gauge).}
\end{equation}
This constrained superfield can be realized as the effective theory of broken supersymmetry where all components but $\phi$ in $\Phi$ acquire mass due to soft supersymmetry breaking.

Let us show the function ${\cal F}_i$ which corresponds to $\mathcal{L}_i$ ($i=2,3,4,5$).
\begin{eqnarray}
\mathcal{F}_2&=&P({\cal X}, \hat{\phi}),\\
\mathcal{F}_3&=&-G_3({\cal X}, \hat{\phi})({\cal D}^a{\cal D}_a\hat{\phi}),\\
\mathcal{F}_4&=&G_{4}({\cal X}, \hat{\phi})\hat{\cal R}_s+G_{4{\cal X}}[({\cal D}^a{\cal D}_a\hat{\phi})^2-({\cal D}_a{\cal D}_b\hat\phi)^2],\\
\mathcal{F}_5&=&G_5({\cal X}, \hat{\phi})\hat{\cal G}^{ab}{\cal D}_a{\cal D}_b\hat{\phi}-\frac16G_{5{\cal X}}({\cal X}, \hat{\phi})[({\cal D}^a{\cal D}_a\hat\phi)^3\nonumber\\
&&-3({\cal D}^a{\cal D}_a\hat\phi)({\cal D}_b{\cal D}_c\hat\phi)^2+2({\cal D}_a{\cal D}_b\hat{\phi})^3]
\end{eqnarray}
where $\hat{\phi}=\frac12 (\Phi+\bar\Phi)$, ${\cal X}=-\frac12 {\cal D}_a\hat\phi{\cal D}^a\hat\phi$, and we have defined
\begin{eqnarray}
({\cal D}_a{\cal D}_b\hat\phi)^2&=&{\cal D}_a{\cal D}_b\hat\phi{\cal D}^a{\cal D}^b\hat\phi,\\
({\cal D}_a{\cal D}_b\hat\phi)^3&=&{\cal D}_a{\cal D}_b\hat\phi{\cal D}^b{\cal D}^c\hat\phi{\cal D}_c{\cal D}^a\hat\phi.
\end{eqnarray}
The superfields $\hat{\cal R}_s$ and ${\cal G}^{ab}$ are given by (see \cite{Farakos:2017mwd})
\begin{eqnarray}
\hat{\cal R}_s&=&-3(\bar{\cal D}_{\dot\alpha}\bar{\cal D}^{\dot\alpha}-8{\cal R})\bar{\cal R}+\frac23 {\cal R}\bar{\cal R}+6G_aG^a-2{\rm i}{\cal D}_aG^a\nonumber\\
\\
\hat{\cal G}^{ab}&=& -\bar{\sigma}^{\dot\alpha \alpha}_b{\cal D}_\alpha\bar{\cal D}_{\dot\alpha}G_a+\left(\frac16\hat{\cal R}_s+4{\cal R}\bar{\cal R}+G_cG^c\right)\eta_{ab}\nonumber\\
&&+2{\rm i}{\cal D}_bG_a+2G_aG_b+\varepsilon^{cd}{}_{ab}{\cal D}_cG_d
\end{eqnarray}
The lowest components of the superfields are as follows (in unitary gauge).
\begin{eqnarray}
\hat{\phi}|&=&\phi,\\
{\cal D}_a \hat{\phi}|&=&e_a^m\partial_m\phi,\\
{\cal D}_a{\cal D}_b \hat{\phi}|&=&e_a^m\nabla_m (e_b^n\partial_n\phi),\\
{\cal R}|&=&-\frac16 M,\\
G_a|&=&-\frac13b_a,\\
\hat{R}_s|&=&\hat{R},\\
\hat{\cal G}|&=&e^a_me^b_n\hat{G}^{mn}.
\end{eqnarray}
Here we emphasize that $\hat{R}$ and $\hat{G}^{mn}$ are Ricci scalar and Einstein tensor consisting of super-covariantized spin-connection $\omega_{m}^{ab}$
\begin{eqnarray}\label{connection}
\omega_m^{ab}&=&-e^{ka}\partial_{[m}e_{k]}^b-e^{lb}\partial_{[l}e_{m]}^a+e^{ka}e^{lb}e_{mc}\partial_{[k}e_{l]}^c\nonumber\\
&&-\frac{\rm i}{2}\psi_m\sigma^{[a}\bar{\psi}^{b]}-\frac{\rm i}{2}\psi^{[a}\sigma^{b]}\bar{\psi}_m-\frac{\rm i}{2}\psi^{[a}\sigma_m\bar{\psi}^{b]},
\end{eqnarray}
where $A_{[a}B_{b]}=\frac12 (A_aB_b-A_bB_a)$. The second order derivative term ${\cal D}_a{\cal D}_b \hat{\phi}|$ also contains the spin-connection.  Thus, we find that these functions ${\cal F}_i$ give the Horndeski Lagrangian for $\phi$ nontrivially coupled to gravitino $\psi_m^\alpha$. Let us show the Horndeski Lagrangian in supergravity:
\begin{eqnarray}
\label{SH}
\mathcal{L}=&&\int d^4\theta E \frac{16 S\bar{S}}{{\cal D}^\alpha{\cal D}_\alpha S\bar{\cal D}_{\dot\alpha}\bar{\cal D}^{\dot\alpha}\bar{S}} \sum_{i=2}^5 {\cal F}_i\nonumber\\
=&&P(X,\phi)-G_3(X,\phi)\nabla^2\phi+G_4(X,\phi)\hat{R}\nonumber\\
&&+G_{4X}[(\nabla^2\phi)^2-({\nabla}_m{\nabla}_n\phi)^2]+G_5(X,\phi)\hat{G}^{mn}{\nabla}_m{\nabla}_n\phi\nonumber\\
&&-\frac16 G_{5X}[(\nabla^2\phi)^3-3(\nabla^2\phi)({\nabla}_m\nabla_n\phi)^2+2({\nabla}_m{\nabla}_n\phi)^3],\nonumber\\
\end{eqnarray}
where we have defined $X=-\frac12\nabla_m\phi\nabla^m\phi$, $(\nabla_m\nabla_n\phi)^2=\nabla_m\nabla_n\phi\nabla^m\nabla^n\phi$ and 
$(\nabla_m\nabla_n\phi)^3=\nabla_m\nabla_n\phi\nabla^n\nabla^k\phi\nabla_k\nabla^m\phi.$
The purely bosonic part of this action is precisely the same as Horndeski model. 

Although the original Horndeski action is known to be free from ghosts,
the absence of the ghosts in our supersymmetric system is still
nontrivial.  Since the gravitino couples to Horndeski Lagrangian through
$\hat{R}$ and $\hat{G}$ as well as  second-order derivative terms of
$\phi$, such as $\nabla^2\phi$, 
one may wonder if the higher-derivative terms for gravitinos are
produced. 
However, we can prove the absence of ghosts as follows.  For the
equations of motion of $\phi$ and graviton, it is proven by a simple
observation that the difference appears only in the explicit form of the
spin-connection. Therefore, the equations of motion of our model are
derived from that of the original Horndeski system by simply replacing
the non-supersymmetric spin-connection by~(\ref{connection}). One finds
that there is no second order derivative term of spin-connection, which
implies that the equation of motion of the gravitino is the first
order. 
Also, our action does not lead to more-than second-order
derivatives for bosons. The gravitino variation in our action is always
associated with the spin-connection. As the third-order derivative terms
are canceled in the equations of motion of $\phi$ and graviton $e_m^a$,
one can check that the coefficients of the variation
$\delta\omega_m{}^{ab}$ is up to the second order in derivatives. 
Since there is no derivative terms of the gravitino $\psi_m^\alpha$ in
the definition of the spin-connection~(\ref{connection}),
$\frac{\delta\omega_m{}^{ab}}{\delta\psi_n^{\alpha}}$ would not increase
the number of derivatives. 
Equivalently, the gravitino equation of motion is up to the first order differential equation of fermion and up to the second order differential equation of bosons. Thus, we find that our supersymmetric Horndeski action~(\ref{SH}) is free from ghosts.

Besides the pure de Sitter supergravity action~(\ref{PSG}), one may add the K\"ahler and superpotential terms of $\Phi$, but then one should take into account the fact that $\Phi$ is a constrained superfield.

We note that the nonlinear realization is supposed to be the LEFT of spontaneously broken supersymmetry, and the theory breaks down at sufficiently high-energy scale. In our case, the cut-off scale is given by the unitarity bound $E_{\rm cut}= \sqrt{6\sqrt{2\pi}m_{3/2}M_{pl}}$, at which four-fermi coupling becomes strong~\cite{Casalbuoni:1988sx}. Also, if gravitino mass is much below the cut-off scale, huge gravitino production may take place~\cite{Hasegawa:2017hgd,Hasegawa:2017nks}. If we consider inflationary models based on our supersymmetric Horndeski Lagrangian, the Hubble scale during inflation should be below the cut-off scale. Such constraint gives the inequality $m_{3/2}>2.8\times10^3 {\rm TeV} \left(\frac{H}{10^{13}{\rm GeV}}\right) $. Therefore, light superparticles would not be expected in our construction, which is very consistent with the collider experimental result so far. 

In this work, we have shown the embedding of Horndeski Lagrangian within (nonlinearly realized) 4D ${\cal N}=1$ supergravity. Supersymmetric higher-derivative couplings often have the issues of ghosts or dynamical auxiliary fields, which have been an obstacle for the construction of general inflationary models with derivative interactions. As we have shown, such issues could be circumvented once supersymmetry is spontaneously broken, and indeed, we have constructed the Horndeski Lagrangian with constrained superfields, which describe the low energy effective theory of broken supersymmetry. In pure de Sitter supergravity model, the physical degrees of freedom are graviton, massive gravitino, and we have added a single real scalar field $\phi$ in the Horndeski Lagrangian. The supergravity realization of Horndeski Lagrangian is achieved in~(\ref{SH}) with a constrained superfield $\Phi$. It is remarkable that gravitino couples to the Horndeski scalar $\phi$ in super-covariant way. As we discussed, additional gravitino couplings do not lead to higher-derivative terms, and do not spoil the ghost-free property of the original Horndeski Lagrangian, although we naively expect fermionic ghosts at first glance. This might be understood as the consequence of supersymmetry of the gravity multiplet: Since graviton is coupled to a real scalar without ghosts, its fermionic partner, gravitino, is also free from ghosts.

We are now able to construct supergravity inflation model with Horndeski Lagrangian, where the inflaton $\phi$ is super-gravitationally coupled to gravitino. In such a case, one needs to consider e.g. the gravitino production through (non-)perturbative processes, which would be calculable on the bases of the Lagrangian we have shown.

It would also be possible to realize larger class of the ghost free scalar-tensor system~\cite{Langlois:2015cwa} within our framework. However, as in our case, additional gravitino couplings would show up and one needs to check whether or not such couplings lead to fermionic ghost instabilities.

\begin{acknowledgments}
YY would like to thank  Hiroaki Tahara for useful discussion and
 Shuntaro Aoki for useful discussion and collaboration on related
 works. The work of YY is supported by Grant-in-Aid for JSPS Fellows
 (19J00494).
JY was supported by JSPS KAKENHI, Grant-in-Aid for Scientific Research 15H02082 and Grant-in-Aid for Scientific Research on Innovative Areas 15H05888.
\end{acknowledgments}


\begin{thebibliography}{99}
\bibitem{Sato:2015dga} For a review of inflation, see, e.g. 
  K.~Sato and J.~Yokoyama,
  Int.\ J.\ Mod.\ Phys.\ D {\bf 24}, no. 11, 1530025 (2015).
\bibitem{Ade:2018gkx} 
  P.~A.~R.~Ade {\it et al.} [BICEP2 and Keck Array Collaborations],
  Phys.\ Rev.\ Lett.\  {\bf 121}, 221301 (2018)
  doi:10.1103/PhysRevLett.121.221301
  [arXiv:1810.05216 [astro-ph.CO]].
  
\bibitem{Bergshoeff:2015tra} 
  E.~A.~Bergshoeff, D.~Z.~Freedman, R.~Kallosh and A.~Van Proeyen,
  Phys.\ Rev.\ D {\bf 92}, no. 8, 085040 (2015)
  Erratum: [Phys.\ Rev.\ D {\bf 93}, no. 6, 069901 (2016)]
  [arXiv:1507.08264 [hep-th]].


\bibitem{Hasegawa:2015bza} 
  F.~Hasegawa and Y.~Yamada,
  JHEP {\bf 1510}, 106 (2015)
  [arXiv:1507.08619 [hep-th]].


\bibitem{Rocek:1978nb} 
  M.~Rocek,
  Phys.\ Rev.\ Lett.\  {\bf 41}, 451 (1978).


\bibitem{Ivanov:1978mx} 
  E.~A.~Ivanov and A.~A.~Kapustnikov,
  J.\ Phys.\ A {\bf 11}, 2375 (1978).


\bibitem{Lindstrom:1979kq} 
  U.~Lindstrom and M.~Rocek,
  Phys.\ Rev.\ D {\bf 19}, 2300 (1979).


\bibitem{Volkov:1973ix} 
  D.~V.~Volkov and V.~P.~Akulov,
  Phys.\ Lett.\  {\bf 46B}, 109 (1973).


\bibitem{Kallosh:2014wsa} 
  R.~Kallosh and T.~Wrase,
  JHEP {\bf 1412}, 117 (2014)
  [arXiv:1411.1121 [hep-th]].


\bibitem{Bergshoeff:2015jxa} 
  E.~A.~Bergshoeff, K.~Dasgupta, R.~Kallosh, A.~Van Proeyen and T.~Wrase,
  JHEP {\bf 1505}, 058 (2015)
  [arXiv:1502.07627 [hep-th]].


\bibitem{Kallosh:2015nia} 
  R.~Kallosh, F.~Quevedo and A.~M.~Uranga,
  JHEP {\bf 1512}, 039 (2015)
  [arXiv:1507.07556 [hep-th]].


\bibitem{Garcia-Etxebarria:2015lif} 
  I.~Garc{\'i}a-Etxebarria, F.~Quevedo and R.~Valandro,
  JHEP {\bf 1602}, 148 (2016)
  [arXiv:1512.06926 [hep-th]].

\bibitem{Cribiori:2019hod} 
  N.~Cribiori, C.~Roupec, T.~Wrase and Y.~Yamada,
  arXiv:1906.07727 [hep-th].
\bibitem{Antoniadis:2014oya} 
  I.~Antoniadis, E.~Dudas, S.~Ferrara and A.~Sagnotti,
  Phys.\ Lett.\ B {\bf 733}, 32 (2014)
  [arXiv:1403.3269 [hep-th]].


\bibitem{Ferrara:2014kva} 
  S.~Ferrara, R.~Kallosh and A.~Linde,
  JHEP {\bf 1410}, 143 (2014)
  [arXiv:1408.4096 [hep-th]].


\bibitem{DallAgata:2014qsj} 
  G.~Dall'Agata and F.~Zwirner,
  JHEP {\bf 1412}, 172 (2014)
  [arXiv:1411.2605 [hep-th]].


\bibitem{McDonough:2016der} 
  E.~McDonough and M.~Scalisi,
  JCAP {\bf 1611}, no. 11, 028 (2016)
  [arXiv:1609.00364 [hep-th]].


\bibitem{Kallosh:2017wnt} 
  R.~Kallosh, A.~Linde, D.~Roest and Y.~Yamada,
  JHEP {\bf 1707}, 057 (2017)
  [arXiv:1705.09247 [hep-th]].

\bibitem{Starobinsky:1980te} 
  A.~A.~Starobinsky,
  Phys.\ Lett.\ B {\bf 91}, 99 (1980)
\bibitem{ArmendarizPicon:1999rj} 
  C.~Armendariz-Picon, T.~Damour and V.~F.~Mukhanov,
  Phys.\ Lett.\ B {\bf 458}, 209 (1999)
  [hep-th/9904075].

\bibitem{Akrami:2018odb} 
  Y.~Akrami {\it et al.} [Planck Collaboration],
  arXiv:1807.06211 [astro-ph.CO].


\bibitem{Kobayashi:2010cm} 
  T.~Kobayashi, M.~Yamaguchi and J.~Yokoyama,
  Phys.\ Rev.\ Lett.\  {\bf 105}, 231302 (2010)
  [arXiv:1008.0603 [hep-th]].
\bibitem{Kobayashi:2011nu} 
  T.~Kobayashi, M.~Yamaguchi and J.~Yokoyama,
  Prog.\ Theor.\ Phys.\  {\bf 126}, 511 (2011)
  [arXiv:1105.5723 [hep-th]].
  
\bibitem{Spokoiny:1984bd} 
  B.~L.~Spokoiny,
  Phys.\ Lett.\  {\bf 147B}, 39 (1984).
  doi:10.1016/0370-2693(84)90587-2
\bibitem{Germani:2010gm} 
  C.~Germani and A.~Kehagias,
  Phys.\ Rev.\ Lett.\  {\bf 105}, 011302 (2010)
  doi:10.1103/PhysRevLett.105.011302
  [arXiv:1003.2635 [hep-ph]].

\bibitem{Horndeski:1974wa} 
  G.~W.~Horndeski,
  Int.\ J.\ Theor.\ Phys.\  {\bf 10}, 363 (1974).

\bibitem{Deffayet:2011gz} 
  C.~Deffayet, X.~Gao, D.~A.~Steer and G.~Zahariade,
  Phys.\ Rev.\ D {\bf 84}, 064039 (2011)
  [arXiv:1103.3260 [hep-th]].
  
\bibitem{Carroll:2003st} 
  S.~M.~Carroll, M.~Hoffman and M.~Trodden,
  Phys.\ Rev.\ D {\bf 68}, 023509 (2003)
  [astro-ph/0301273].
  
\bibitem{Woodard:2006nt} 
  R.~P.~Woodard,
  Lect.\ Notes Phys.\  {\bf 720}, 403 (2007)
  [astro-ph/0601672].


\bibitem{Sbisa:2014pzo} 
  F.~Sbis\`a,
  Eur.\ J.\ Phys.\  {\bf 36}, 015009 (2015)
  [arXiv:1406.4550 [hep-th]].

\bibitem{Aoki:2018lwx} 
  K.~Aoki and K.~Shimada,
  Phys.\ Rev.\ D {\bf 98}, no. 4, 044038 (2018)
  [arXiv:1806.02589 [gr-qc]].
  
\bibitem{Aoki:2019rvi} 
  K.~Aoki and K.~Shimada,
  arXiv:1904.10175 [hep-th].
\bibitem{Baker:2017hug} 
T.~Baker, E.~Bellini, P.~G.~Ferreira, M.~Lagos, J.~Noller and I.~Sawicki,
Phys.\ Rev.\ Lett.\ {\bf 119}, no. 25, 251301 (2017)
[arXiv:1710.06394 [astro-ph.CO]].
\bibitem{Creminelli:2017sry} 
P.~Creminelli and F.~Vernizzi,
Phys.\ Rev.\ Lett.\ {\bf 119}, no. 25, 251302 (2017)
[arXiv:1710.05877 [astro-ph.CO]].
\bibitem{Sakstein:2017xjx} 
J.~Sakstein and B.~Jain,
Phys.\ Rev.\ Lett.\ {\bf 119}, no. 25, 251303 (2017)
[arXiv:1710.05893 [astro-ph.CO]].
\bibitem{Ezquiaga:2017ekz} 
J.~M.~Ezquiaga and M.~Zumalacarregui,
Phys.\ Rev.\ Lett.\ {\bf 119}, no. 25, 251304 (2017)
[arXiv:1710.05901 [astro-ph.CO]].
 %

\bibitem{Copeland:2018yuh} 
  E.~J.~Copeland, M.~Kopp, A.~Padilla, P.~M.~Saffin and C.~Skordis,
  Phys.\ Rev.\ Lett.\  {\bf 122}, no. 6, 061301 (2019)
  [arXiv:1810.08239 [gr-qc]].


\bibitem{Kobayashi:2019hrl} 
T.~Kobayashi,
arXiv:1901.07183 [gr-qc].  To be published in Rep.\ Prog.\ Phys.


\bibitem{Khoury:2010gb} 
  J.~Khoury, J.~L.~Lehners and B.~Ovrut,
  Phys.\ Rev.\ D {\bf 83}, 125031 (2011)
  [arXiv:1012.3748 [hep-th]].


\bibitem{Khoury:2011da} 
  J.~Khoury, J.~L.~Lehners and B.~A.~Ovrut,
  Phys.\ Rev.\ D {\bf 84}, 043521 (2011)
  [arXiv:1103.0003 [hep-th]].


\bibitem{Farakos:2012je} 
  F.~Farakos, C.~Germani, A.~Kehagias and E.~N.~Saridakis,
  JHEP {\bf 1205}, 050 (2012)
  [arXiv:1202.3780 [hep-th]].


\bibitem{Sasaki:2012ka} 
  S.~Sasaki, M.~Yamaguchi and D.~Yokoyama,
  Phys.\ Lett.\ B {\bf 718}, 1 (2012)
  [arXiv:1205.1353 [hep-th]].


\bibitem{Koehn:2012ar} 
  M.~Koehn, J.~L.~Lehners and B.~A.~Ovrut,
  Phys.\ Rev.\ D {\bf 86}, 085019 (2012)
  [arXiv:1207.3798 [hep-th]].


\bibitem{Koehn:2012np} 
  M.~Koehn, J.~L.~Lehners and B.~A.~Ovrut,
  Phys.\ Rev.\ D {\bf 86}, 123510 (2012)
  [arXiv:1208.0752 [hep-th]].


\bibitem{Farakos:2012qu} 
  F.~Farakos and A.~Kehagias,
  JHEP {\bf 1211}, 077 (2012)
  [arXiv:1207.4767 [hep-th]].


\bibitem{Farakos:2013zya} 
  F.~Farakos, C.~Germani and A.~Kehagias,
  JHEP {\bf 1311}, 045 (2013)
  [arXiv:1306.2961 [hep-th]].


\bibitem{Gwyn:2014wna} 
  R.~Gwyn and J.~L.~Lehners,
  JHEP {\bf 1405}, 050 (2014)
  [arXiv:1402.5120 [hep-th]].


\bibitem{Aoki:2014pna} 
  S.~Aoki and Y.~Yamada,
  Phys.\ Rev.\ D {\bf 90}, no. 12, 127701 (2014)
  [arXiv:1409.4183 [hep-th]].


\bibitem{Aoki:2015eba} 
  S.~Aoki and Y.~Yamada,
  JCAP {\bf 1507}, no. 07, 020 (2015)
  [arXiv:1504.07023 [hep-th]].


\bibitem{Fujimori:2016udq} 
  T.~Fujimori, M.~Nitta and Y.~Yamada,
  JHEP {\bf 1609}, 106 (2016)
  [arXiv:1608.01843 [hep-th]].


\bibitem{Fujimori:2017kyi} 
  T.~Fujimori, M.~Nitta, K.~Ohashi, Y.~Yamada and R.~Yokokura,
  JHEP {\bf 1709}, 143 (2017)
  [arXiv:1708.05129 [hep-th]].


\bibitem{Fujimori:2017rcc} 
  T.~Fujimori, M.~Nitta, K.~Ohashi and Y.~Yamada,
  JHEP {\bf 1805}, 102 (2018)
  [arXiv:1712.05017 [hep-th]].


\bibitem{Nitta:2018yzb} 
  M.~Nitta and R.~Yokokura,
  JHEP {\bf 1810}, 146 (2018)
  [arXiv:1809.03957 [hep-th]].


\bibitem{Nitta:2018vyc} 
  M.~Nitta and R.~Yokokura,
  JHEP {\bf 1905}, 102 (2019)
  [arXiv:1810.12678 [hep-th]].
  %
 

\bibitem{Wess:1992cp} 
  J.~Wess and J.~Bagger,
  ``Supersymmetry and supergravity,'' (Princeton University Press, 1992)


\bibitem{Komargodski:2009rz} 
  Z.~Komargodski and N.~Seiberg,
  JHEP {\bf 0909}, 066 (2009)
  [arXiv:0907.2441 [hep-th]].


\bibitem{Farakos:2017mwd} 
  F.~Farakos, S.~Ferrara, A.~Kehagias and D.~L\"ust,
  Fortsch.\ Phys.\  {\bf 65}, no. 12, 1700073 (2017)
  [arXiv:1707.06991 [hep-th]].


\bibitem{Ferrara:2015tyn} 
  S.~Ferrara, R.~Kallosh and J.~Thaler,
  Phys.\ Rev.\ D {\bf 93}, no. 4, 043516 (2016)
  [arXiv:1512.00545 [hep-th]].


\bibitem{Carrasco:2015iij} 
  J.~J.~M.~Carrasco, R.~Kallosh and A.~Linde,
  Phys.\ Rev.\ D {\bf 93}, no. 6, 061301 (2016)
  [arXiv:1512.00546 [hep-th]].


\bibitem{DallAgata:2015zxp} 
  G.~Dall'Agata and F.~Farakos,
  JHEP {\bf 1602}, 101 (2016)
  [arXiv:1512.02158 [hep-th]].


\bibitem{Ferrara:2016een} 
  S.~Ferrara, R.~Kallosh, A.~Van Proeyen and T.~Wrase,
  JHEP {\bf 1604}, 065 (2016)
  [arXiv:1603.02653 [hep-th]].

\bibitem{Casalbuoni:1988sx} 
  R.~Casalbuoni, S.~De Curtis, D.~Dominici, F.~Feruglio and R.~Gatto,
  Phys.\ Lett.\ B {\bf 216}, 325 (1989)
  Erratum: [Phys.\ Lett.\ B {\bf 229}, 439 (1989)].
  
\bibitem{Hasegawa:2017hgd} 
  F.~Hasegawa, K.~Mukaida, K.~Nakayama, T.~Terada and Y.~Yamada,
  Phys.\ Lett.\ B {\bf 767}, 392 (2017)
  [arXiv:1701.03106 [hep-ph]].

\bibitem{Hasegawa:2017nks} 
  F.~Hasegawa, K.~Nakayama, T.~Terada and Y.~Yamada,
  Phys.\ Lett.\ B {\bf 777}, 270 (2018)
  [arXiv:1709.01246 [hep-ph]].
  
\bibitem{Langlois:2015cwa} 
  D.~Langlois and K.~Noui,
  JCAP {\bf 1602}, no. 02, 034 (2016)
  [arXiv:1510.06930 [gr-qc]].
\end{thebibliography}
\end{document}